%% file: main.tex
\newif\ifdouble
\newif\ifsingle
\newif\ifchange
\doubletrue

  \documentclass[sigconf]{acmart}

\usepackage{dblfloatfix} 
\usepackage{booktabs} 
\usepackage{arydshln}
\usepackage{subfig}
\usepackage{amsmath}
\AtBeginDocument{%
  \providecommand\BibTeX{{%
    \normalfont B\kern-0.5em{\scshape i\kern-0.25em b}\kern-0.8em\TeX}}}

\copyrightyear{2022}
\acmYear{2022}
\setcopyright{rightsretained}
\acmConference[UIST '22 Adjunct]{The Adjunct Publication of the 35th
Annual ACM Symposium on User Interface Software and
Technology}{October 29-November 2, 2022}{Bend, OR, USA}
\acmBooktitle{The Adjunct Publication of the 35th Annual ACM
Symposium on User Interface Software and Technology (UIST '22 Adjunct),
October 29-November 2, 2022, Bend, OR, USA}
\acmDOI{10.1145/3526114.3558736}
\acmISBN{978-1-4503-9321-8/22/10}

\newcommand{\system}{HapticLever}

\begin{document}
\pagenumbering{arabic}
\pagestyle{plain}
\title{\system{}: Kinematic Force Feedback using a 3D Pantograph}

\author{Marcus Friedel}
\affiliation{%
  \institution{University of Calgary}
  \city{Calgary}
  \country{Canada}}
\email{marcus.friedel@ucalgary.ca}

\author{Ehud Sharlin}
\affiliation{%
  \institution{University of Calgary}
  \city{Calgary}
  \country{Canada}}
\email{ehud@cpsc.ucalgary.ca}

\author{Ryo Suzuki}
\affiliation{%
  \institution{University of Calgary}
  \city{Calgary}
  \country{Canada}}
\email{ryo.suzuki@ucalgary.ca}

\renewcommand{\shortauthors}{Friedel, et al.}
\input{0-abstract}

\begin{teaserfigure}
\includegraphics[width=\textwidth]{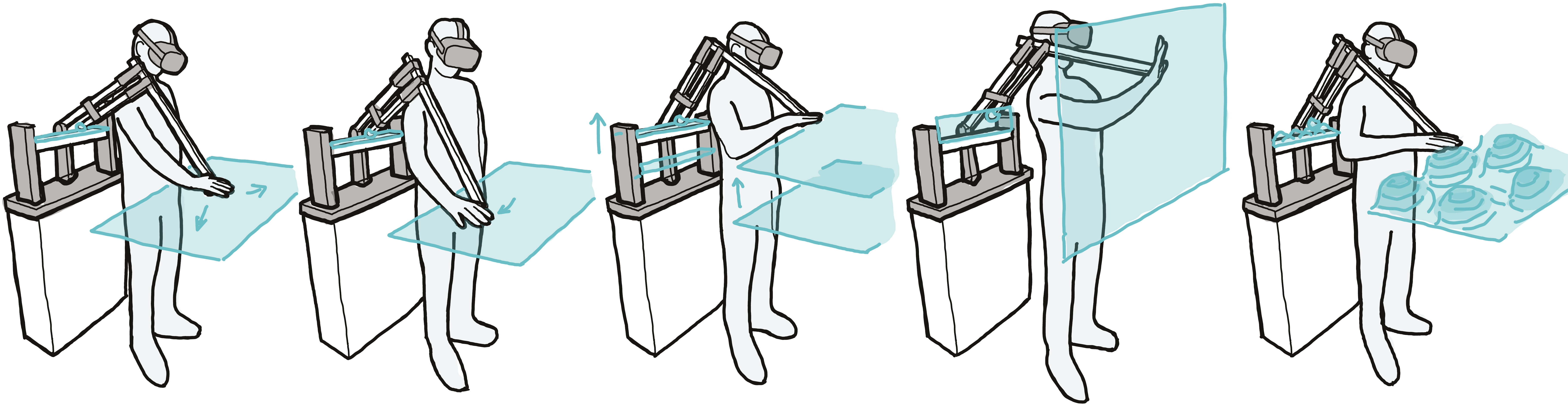}
\caption{\system{} is a design concept for large-scale VR haptics which passively transforms a small-scale constraint into a stiff, realistic, large-scale render.}
\Description{A series of illustrations showing HapticLever use cases, including: exploring a horizontal flat virtual surface, hitting a virtual surface, interacting with a height-changing virtual surface, exploring a vertical flat virtual surface, and exploring a surface with an irregular shape.}
\label{fig:teaser}
\end{teaserfigure}

\maketitle

\input{1-introduction}

\input{3-concept}

\input{7-future-work}

\ifdouble
  \balance
\fi
\bibliographystyle{ACM-Reference-Format}
\bibliography{references}

\input{9-appendix}
\end{document}

%% file: 0-abstract.tex
\begin{abstract}
\system{} is a new kinematic approach for VR haptics which uses a 3D pantograph to stiffly render large-scale surfaces using small-scale proxies.
The \system{} approach does not consume power to render forces, but rather puts a mechanical constraint on the end effector using a small-scale proxy surface.
The \system{} approach provides stiff force feedback when the user interacts with a static virtual surface, but allows the user to move their arm freely when moving through free virtual space.
We present the problem space, the related work, and the \system{} design approach.
\end{abstract}

\ccsdesc[500]{Human-centered computing~Virtual reality}
\ccsdesc[500]{Human-centered computing~Haptic devices}

\keywords{haptics, force feedback, virtual reality, robotics, user interfaces}


%% file: 1-introduction.tex
\section{Introduction and Related Work}

Rendering large-scale, high-stiffness shapes and surfaces remains a challenge in haptics.
Most of today's haptic devices are handheld \cite{shigeyama2019transcalibur, Ryu2021GamesBond, Gonzalez2021XRings} or focus on hand and finger haptics \cite{Sinclair2019Capstan, benko2016normaltouch, choi2016wolverine, choi2018claw}. 
Those which attempt to render a net force on the user \cite{heo2018thor, choi2017grabity, Lopes2017HapticsToWallsByEMS} are unable to render stiff interactions. 
Haptic devices designed for large-scale interactions either use a force-based dynamics approach or a force-independent kinematics approach. 
Dynamics approaches to haptics are concerned with the forces on the user, whereas Kinematics approaches are concerned with the user's allowed motion. 
Large-scale, high-stiffness physical interactions such as a hand on a tabletop can be better represented as a degree of freedom reduction of the hand, rather than as a time-series of applied forces. 

Large-scale dynamics approaches \cite{Barnaby2019Mantis, vonach2017vrrobot, calvo} shake and oscillate when rendering stiff surfaces, must consumer power to render any force, and encumber the user with a constant resistance. 
These devices shake and oscillate because of the cantilever effect on their serial links and because they use a feedback control system which interfaces with the unpredictable, unmodellable human user. 
The maximum force these devices can apply is limited by the actuator strength, and if users exceed this limit then the device will oscillate or break.
Because these devices directly actuate the mechanism joints directly, they must consume power to engage a force, to sustain a force, and often to compensate for gravity even when not rendering a force. 
These devices constantly resist the user's motion because of acturator back-torque, in the case of impedance control, or active motor pushback, in the case of admittance control.
The advantages of the dynamics approach are that it needs only to provide an on-demand touchpoint for the user and that it is versatile and can be programmed to render many different interactions.

Kinematics approaches primarily take the form of reconfigurable proxies at large-scale and mechanical constraints at small-scale. 
These function as feed-forward positional control, in contrast to the force-in-the-loop feedback control of dynamics systems.
Large-scale approaches \cite{suzuki2020roomshift, yixian2020zoomwalls, wang2020movevr, Cheng2015TurkDeck, Cheng2014HapticTurk, suzuki2020lifttiles, Teng2019TilePop, suzuki2022augmented} aim to reconfigure props and proxies in the physical environment in order to provide haptic sensations.
The user is free to engage with or disengage from the touchpoint, and the control system is only concerned with representing the virtual objects, rather than reacting to the user. 
While interacting with full-sized proxies is realistic and can provide stiff interactions, they have high space requirements.
Small-scale kinematics approaches \cite{wireality2020, Hinchet2018DextrES, Li2020HapLinkage, choi2016wolverine, Nith2021DextrEMS} use mechanical constraints to restrict the motion of the user's hands or fingers, giving the user the freedom to apply light, heavy, or variable forces with no displacement. 
While these small-scale approaches can serve as inspiration, their mechanical constraints typically limit the user to zero or one degree of freedom, rather than the three degrees of freedom that a hand on a 2D surface has.

\system{} is an interface which can provide large-scale haptic interactions using a 3D pantograph to scale-up the mechanical constraint of a small-scale proxy.

%% file: 3-concept.tex
\section{Design}

This section introduces a new pantograph-based approach to providing stiff, transparent, large-scale haptic experiences.

The design must:

\begin{itemize}

    \item Provide instant, stiff, passive force feedback when the user interacts with a static virtual surface.

    \item Allow the user to move their arm naturally, without resistance, when moving through free virtual space.

\end{itemize}

\begin{figure}[h]
    \centering
    \includegraphics[width=\columnwidth]{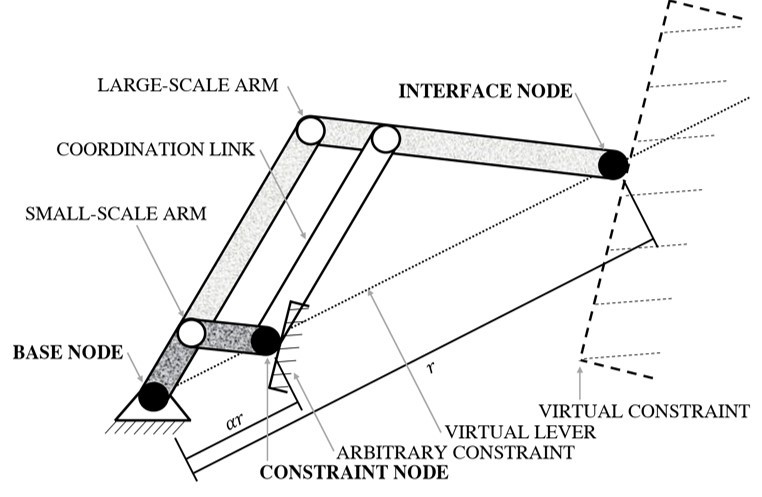}
    \caption{A 2D pantograph with the \system{} labels.}
    \label{fig:LeverEvolution}
    \Description{A diagram of a 2D pantograph labelled with the HapticLever labels.}
\end{figure}

\begin{figure}[h]
    \centering
    \includegraphics[width=\columnwidth]{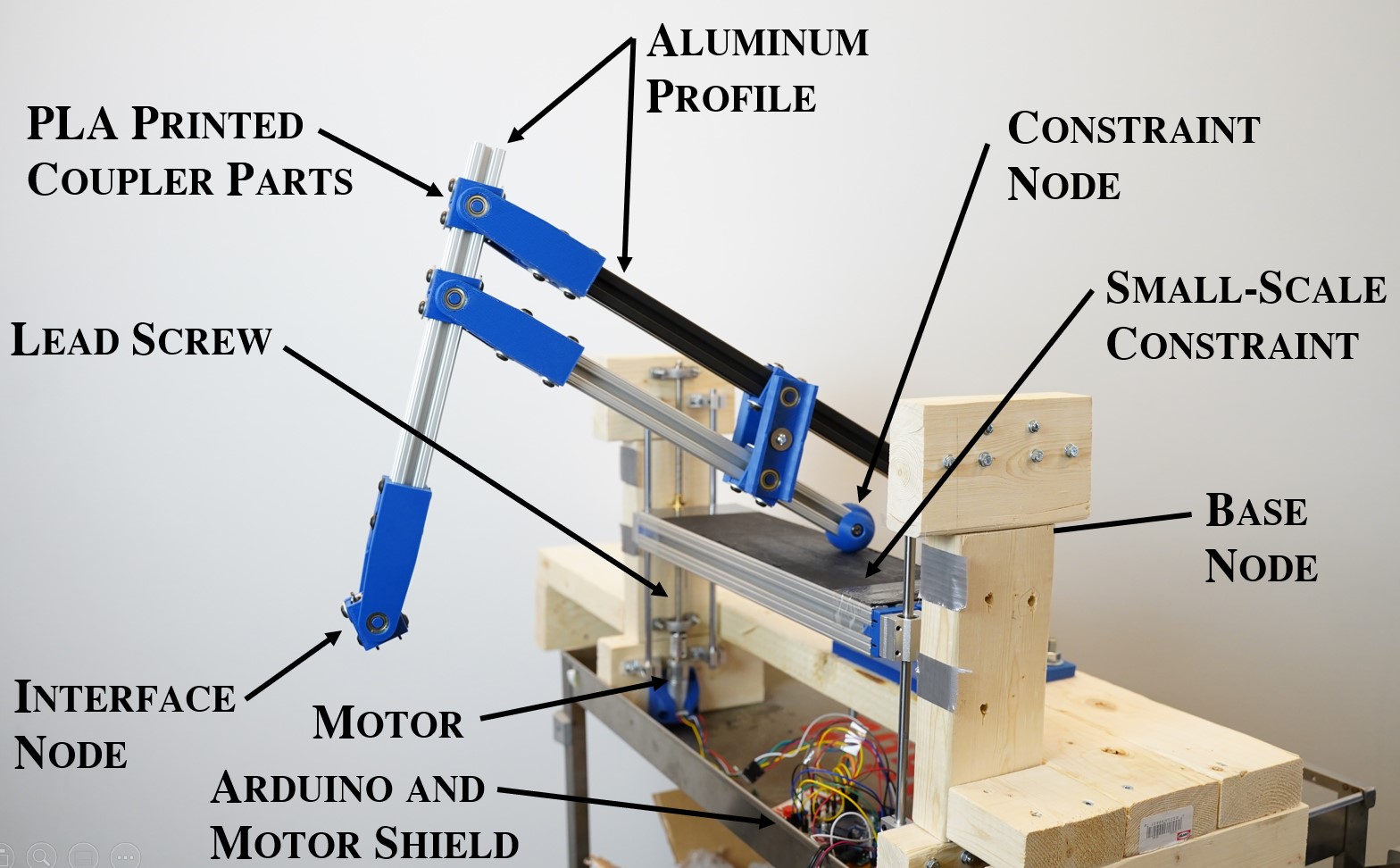}
    \caption{A labelled image of our implementation.}
    \label{fig:Implementation}
    \Description{A labelled photograph of the HapticLever implementation.}
\end{figure}

The pantograph design, shown in Figure \ref{fig:LeverEvolution}, acts as a virtual lever in which the load (constraint) and effort (interface) nodes can change their absolute radial position, but remain at proportionally constant radial positions relative to one another. The pantograph design means that the constraint and interface nodes follow scaled paths, and therefore share scaled positional constraints. \system{}, shown in Figure \ref{fig:Implementation}, extends the pantograph mechanism into 3D by adding a vertical rotational joint at the base node. 

The primary tradeoff is the advantageously smaller workspace of the constraint node versus the unfavourably larger forces at the constraint node. The pantograph scaling factor means that a small constraint at the constraint node corresponds to a large constraint at the interface node.

Because they are rigidly connected, any force felt by the constraint node is transferred passively through the links to the interface node. The force tolerance of the system is limited by the mechanical strength of the linkage, rather than the strength of any actuator. Therefore, \system{} can withstand higher forces than the direct actuation counterparts.
If the proxy surface or object is rigidly connected to the base node, therefore forming a rigid loop between the base and constraint nodes, then \system{} does not consume any energy to transform the force from the constraint node to the interface node. 
Therefore, in the current implementation, the base surface underneath the constraint node is moved using self-locking lead screws. 
Because no actuators are directly attached to the pantograph mechanism, the user can transparently move the interface node free of resistance whenever the constraint node is not providing force feedback.

\begin{figure}[h]
    \centering
    \includegraphics[width=\columnwidth]{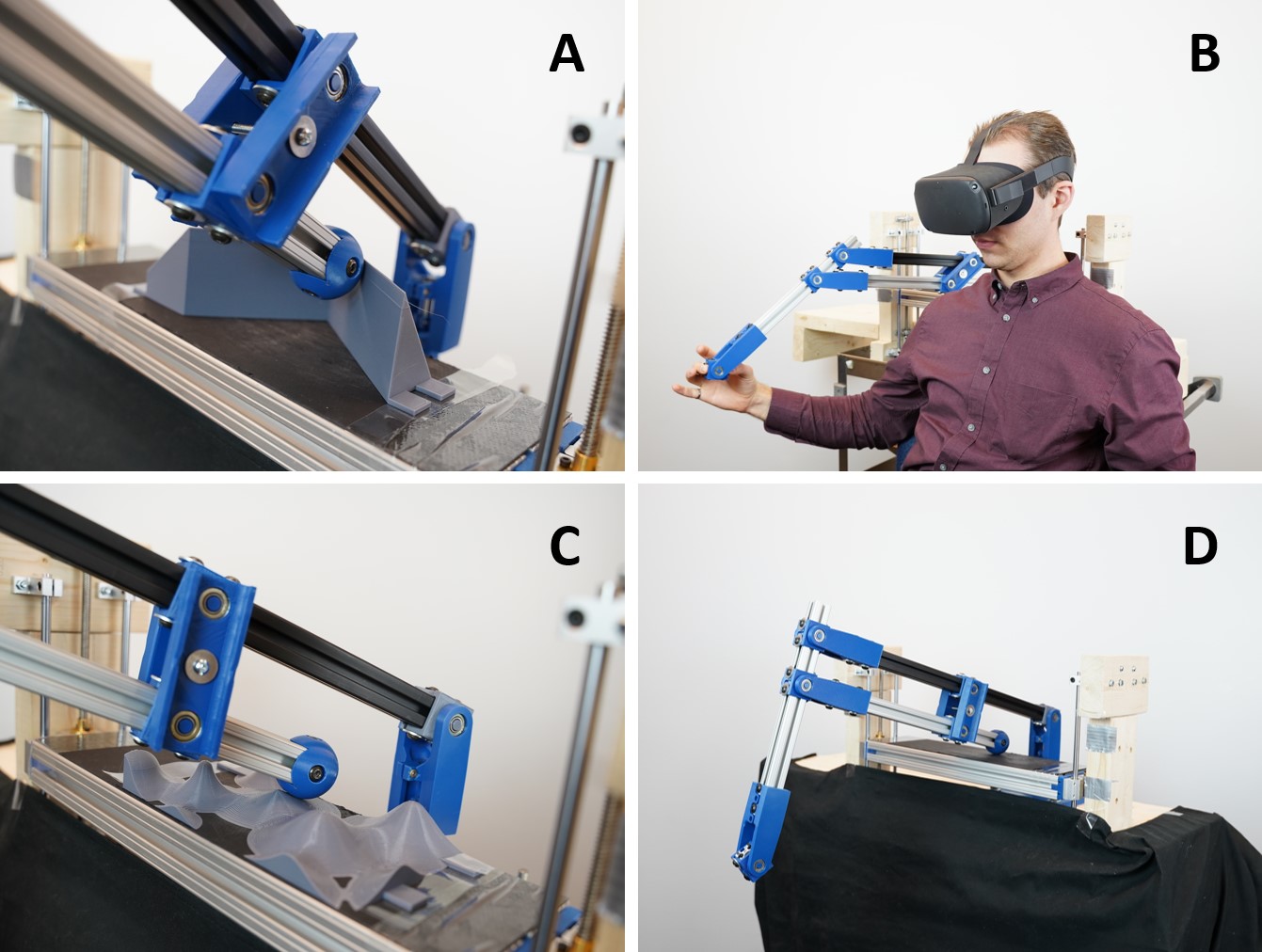}
    \caption{\system{} demonstrations. A: Vertical walls at a corner. B: Drawing on a surface with a stabilized hand. C: An irregular surface. D: A flat horizontal plane.}
    \label{fig:DemoApplications}
    \Description{Four photographs showing demonstrations of the HapticLever implementation: A: Interacting with vertical walls at a corner. B: Drawing on a surface with a stabilized hand. C: Interacting with an irregular surface. D: Interacting with a flat horizontal plane.}
\end{figure}

The demonstrations shown in Figure \ref{fig:DemoApplications} include vertical walls at a corner, drawing on a surface with a stabilized hand, exploring an irregular surface, and a horizontal plane. These different renderings correspond to custom proxy objects. 
Users can press lightly on the rendered surface, press heavily on the rendered surface, slide their hand along the rendered surface, or hit the surface with an impact---and in all cases the surface will respond like a real table.

In the implementation, the weight of the pantograph linkage is 1.68kg, the weight of the entire \system{} system including the base is 9.89kg, and the weight on hand which the user feels is 0.91kg.
The rotatable handle at the interface node gives the system five degrees of freedom.
The workspace of the interface node is a portion of a spherical shell with inner radius 342mm, outer radius 722mm, and solid angle 2.33 steradians. This workspace is large enough to prototype body-scale haptic interactions.
The average force downwards force tolerance on the interface node is 64N, and the average downwards impulse tolerance on the interface node is 7.3kg$\cdot$m/s.

%% file: 7-future-work.tex
\section{Future Work}

\system{} can be developed further in the following ways:
While the current design tranfsers linear force through the pantograph, future work could investigate transferring rotational moments.
A \system{} pantograph is not limited to a base node, one constraint node, and one interface node; adding another parallelogram to the linkage would create a second constraint node. These two constraint nodes could simultaneously render different constraints, such as intersecting walls.
The future work of turning \system{} into a portable, wearable device entails redesigning for weight and comfort, and ensuring that the reaction forces are appropriately grounded on the user's torso.
Replacing the small-scale proxy with actuators or a shape display like shapeShift \cite{Siu2018ShapeShift} would enable \system{} to change the position, orientation, or shape of the constraint. To maintain stiffness via the mechanical constraint, such actuators or shape display should be mechanically self locking.
\system{}  behaves stiffly when encountering the small-scale constraint and moves without resistance when not encountering the small-scale constraint.
\system{} lays a groundwork for promising future large-scale haptic devices that provide stiff and natural interactions.